\documentclass{ws-p8-50x6-00}

\begin{document}

\title{Galactic Chemical Evolution and the abundances of lithium, beryllium
and boron}

\author{Andreu Alib\'es, Javier Labay and Ramon Canal}
\address{Departament d'Astronomia i Meteorologia, Universitat de Barcelona,
Mart\'{\i} i Franqu\`es 1, 08028 Barcelona, Spain.\ E-mail: aalibes, javier,
ramon@am.ub.es}
\maketitle

\abstracts{A LiBeB evolution model including Galactic Cosmic Ray nucleosynthesis, the
$\nu$-process, novae, AGB and C-stars is presented.}

%\section{Introduction. The Galactic Chemical Evolution Model}

We have included Galactic Cosmic Ray Nucleosynthesis (GCRN) in a complete Chemical Evolution Model
that takes into account 76 stable isotopes from hydrogen to zinc. Any successful
LiBeB evolution model should also be compatible with other observational constraints
like the age-metallicity relation, the G-dwarf distribution or the evolution
of other elements.
At the same time, we have checked how different would be a model that took into
account the last observations by Wakker et al. (1999) of metal-enriched clouds
falling onto the disk, from a primordial infall model.
%\section{Galactic Chemical Evolution Model}

We have integrated the standard evolution equations, using a SFR proportional to $\sigma^m \sigma_g ^n$ and a double exponential infall (Chiappini
et al. 1997) with \( \tau _{h} \) = 1 Gyr, \( \tau _{d} \) = 7 Gyr and \( t_{max} \)
= 1 Gyr. We have used two kinds of infall material: in the primordial model
(PM), the galaxy acretes only primordial material during all its lifetime
(13 Gyr), while the enriched model (EM) considers a first Gyr of primordial
infall, followed by 12 Gyr of enriched infall (0.1 Z\( _{\odot } \), indicated
by recent observations of clouds falling onto the disk by Wakker et al. (1999)).

We use the Kroupa et al. (1993) IMF
and different prescriptions for the nucleosynthesis in the various mass ranges:
low and intermediate-mass stars (van den Hoek and Groenewegen, 1997), type II
SNe (Woosley and Weaver, 1995) and type Ia SNe (model W7 of Nomoto et al., 1997).
Finally, novae have also been included, by means of the Jos\'e and Hernanz (1998)
yields.

%\section{LiBeB Model}
\vskip0.5cm

The main source of \( ^{6} \)Li, \( ^{9} \)Be and \( ^{10} \)B is the \textbf{GCRN}. In the superbubble (SB) scenario, we have calculated
the production rate by this mechanism, taking a GCR composition that comes from inside the SB, where newly synthesized material
ejected by a SN is accelerated by the shock wave of other SN and mixed with
the ISM at that epoch. 
The energy spectrum of the GCR is:~\texttt{\( q(E)\propto \frac{p^{-2.2}}{\beta }e^{-\frac{E}{E_{0}}} \)},
with \( E_{0} \) = 10 GeV/n.
For \textbf{the $\nu$-process} we have fixed the contribution of WW95 yields (about
25\%) by means of (\( ^{11} \)B/\( ^{10} \)B)\( _{\odot } \).
\textbf{AGB stars and C-stars} are also producers of \( ^{7} \)Li. We use the
time-dependent production rate suggested by Abia et al. (1993) and a current 
production rate of 1.5\( \cdot  \)10\( ^{-8} \) M\( _{\odot } \)pc\( ^{-2} 
\)Gyr\( ^{-1} \).
Finally, we have taken an average yield per \textbf{nova} \textbf{outburst}
of 1.03\( \cdot  \)10\( ^{-10} \) M\( _{\odot } \), considering that 30\%
of the outbursts come from ONe WD and 70\% from CO ones.

\begin{figure}[t]
%\figurebox{20pc}{15pc}{} % to have a box alone
\epsfxsize=15pc % will enlarge or reduce the postscript figures based on the xsize
\epsfbox{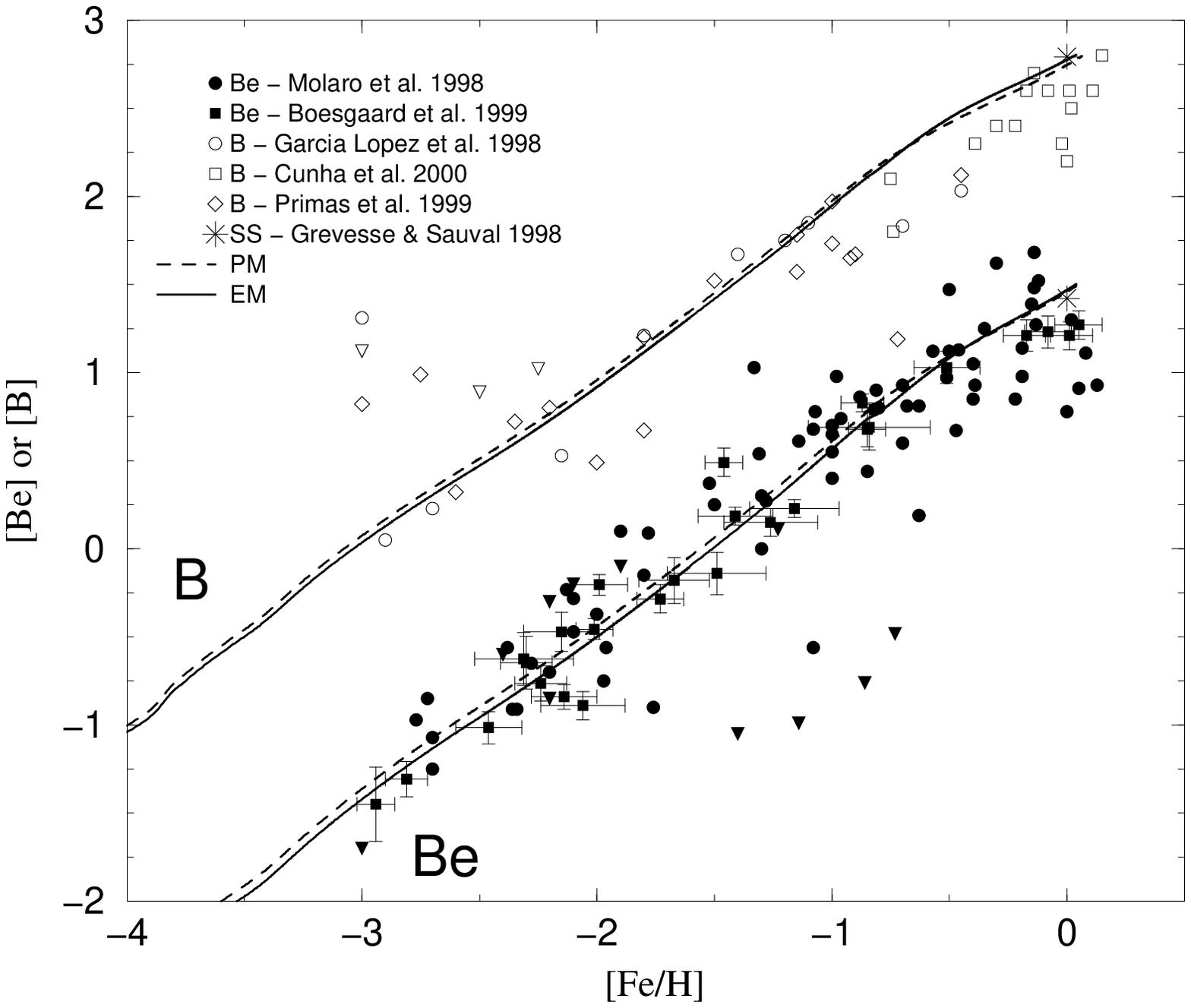} % postscript image file name
\epsfxsize=15pc % will enlarge or reduce the postscript figures based on the xsize
\epsfbox{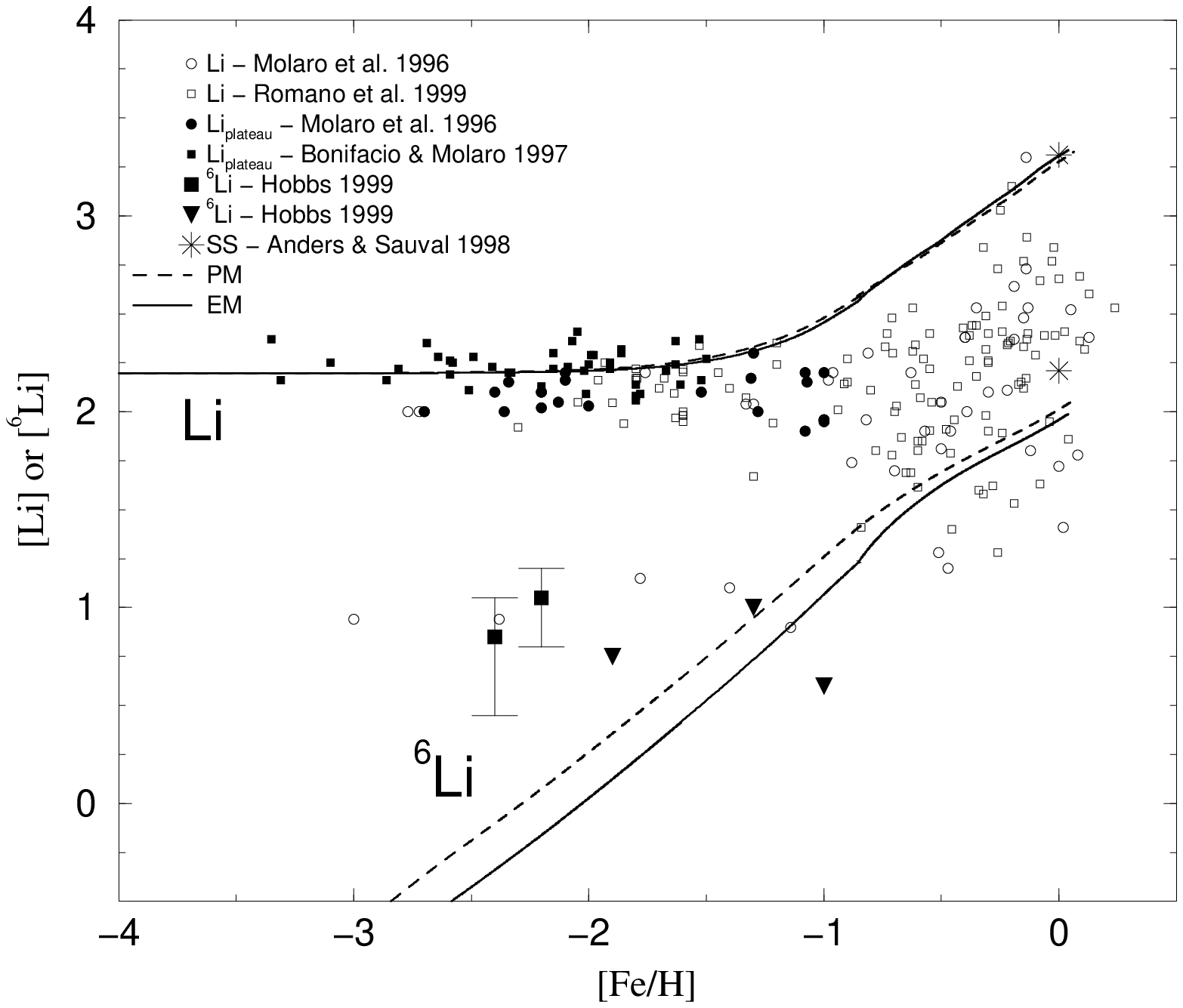} % postscript image file name
%\caption{For the two compositions of the infall a) Beryllium and boron; b) Lithium
%and $^6$Li.}
\end{figure}
%\section{Conclusions}
\vskip0.5cm
Our two-infall model, both primordial
and enriched, is able to reproduce the main solar neighborhood data; in particular,
the G-dwarf distribution, which is the main constraint, and the evolution of
the majority of chemical elements, specially the CNO ones.
Together with the superbubble scenario for the acceleration of
GCR is also able to reproduce the LiBeB evolution, specially the linear relationships
of Be vs Fe and B vs Fe, when taking a GCR composition as a mixture of ejecta
of SNe (20\%) and ISM material (80\%).\\
All the $^7$Li sources considered are necessary to reproduce
the Li evolution.


\begin{thebibliography}{99}
\bibitem{} Abia, C., Isern, J., Canal, R. 1993 A\&A \textbf{275}, 96
\bibitem{} Chiappini, C., Matteucci, F., \& Gratton, R. 1997 ApJ \textbf{477}, 765
\bibitem{} Jos\'e, J., \& Hernanz, M. 1998 ApJ \textbf{494}, 680
\bibitem{} Kroupa, P., Tout, C., \& Gilmore, G. 1993 MNRAS \textbf{262}, 545
\bibitem{} Nomoto, et al. 1997 Nucl. Phys. A \textbf{621}, 467c
\bibitem{} van den Hoek, L.B., \& Groenewegen, M.A.T. 1997 A\&ASS \textbf{123}, 305
\bibitem{} Wakker, B.P. et al. 1999 Nature \textbf{402}, 338
\bibitem{} Woosley, S.E., \& Weaver, T.A. 1995 ApJS \textbf{101}, 181 

\end{thebibliography}
\end{document}